\documentclass[twocolumn,aps,showpacs,prl]{revtex4}
\usepackage{graphicx}
\usepackage{epstopdf}
\usepackage{amsmath}
\usepackage{multirow}

\begin{document}

\title{Electronic structures and magnetic orders of Fe-vacancies ordered ternary
iron selenides TlFe$_{1.5}$Se$_2$ and AFe$_{1.5}$Se$_2$ (A=K, Rb, or Cs)}

\author{Xun-Wang Yan$^{1,2}$}
\author{Miao Gao$^{1}$}
\author{Zhong-Yi Lu$^{1}$}\email{zlu@ruc.edu.cn}
\author{Tao Xiang$^{2,3}$}\email{txiang@iphy.ac.cn}

\date{\today}

\affiliation{$^{1}$Department of Physics, Renmin University of
China, Beijing 100872, China}

\affiliation{$^{2}$Institute of Theoretical Physics, Chinese Academy
of Sciences, Beijing 100190, China }

\affiliation{$^{3}$Institute of Physics, Chinese Academy of
Sciences, Beijing 100190, China }

\begin{abstract}

By the first-principles electronic structure calculations, we find
that the ground state of the Fe-vacancies ordered TlFe$_{1.5}$Se$_2$
is a quasi-two-dimensional collinear antiferromagnetic semiconductor
with an energy gap of 94 meV, in agreement with experimental
measurements. This antiferromagnetic order is driven by the
Se-bridged antiferromagnetic superexchange interactions between Fe
moments. Similarly, we find that crystals AFe$_{1.5}$Se$_2$ (A=K,
Rb, or Cs) are also antiferromagnetic semiconductors but with a
zero-gap semiconducting state or semimetallic state nearly
degenerated with the ground states. Thus rich physical properties
and phase diagrams are expected.

\end{abstract}


\maketitle


The discovery of high transition temperature superconductivity in
LaFeAsO by partial substitution of O with F atoms \cite{kamihara}
stimulates great interest on iron pnictides. Other kinds of
iron-based compounds were also reported to show superconductivity
after doping or under high pressures\cite{rotter,wang,hsu}. A
ubiquitous feature is that the parent compounds of these
superconductors are antiferromagnetic (AFM) semimetals\cite{lu1}
with either a collinear \cite{cruz,dong} or
bi-collinear\cite{ma,bao,shi} AFM order below a structural
transition temperature.

Very recently the superconductivity was discovered at about 30K in
the potassium intercalated FeSe-layer compound K$_{0.8}$Fe$_2$Se$_2$
\cite{chen}. Soon after, the superconductivity was also found in the
Cs-intercalated compound Cs$_{0.8}$(FeSe$_{0.98}$)$_2$ \cite{Cs} and
(Tl,K)-intercalated compound (Tl,K)Fe$_x$Se$_2$\cite{fang}. These
compounds have the ThCr$_{2}$Si$_{2}$ type structure (Fig. 1(a)),
isostructural with 122-type iron pnictides BaFe$_2$As$_2$
\cite{rotter}. But they should be regarded as a new kind of
iron-based superconductors since they are chalcogenides rather than
pnictides. We have done the first-principles electronic structure
calculations for these materials \cite{yan}. We find that their
parent compounds TlFe$_2$Se$_2$ and AFe$_2$Se$_2$  (A=K or Cs) are
also AFM semimetals and the ground states are in a bi-collinear AFM
order.

\begin{figure}
\includegraphics[width=8cm]{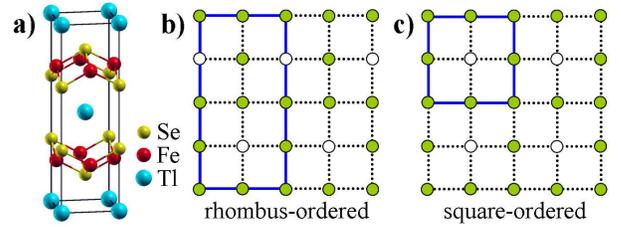}
\caption{(Color online) TlFe$_x$Se$_2$ with the ZrCuSiAs-type
structure: (a) a tetragonal unit cell containing two formula units
without any vacancy; (b) schematic top view of the Fe-Fe square
layer with one quarter Fe-vacancies ordered in rhombus ($x=1.5$), in
which there are two inequivalent Fe atoms, namely 2-Fe-neighbored
and 3-Fe-neighbored Fe atoms. (c) schematic top view of the Fe-Fe
square layer with one quarter Fe-vacancies ordered in square
($x=1.5$). The filled circles denote the Fe atoms while the empty
circles denote the Fe-vacancies. The square and rectangle enclosed
by the solid lines denote the unit cells.} \label{figa}
\end{figure}

An important feature revealed by the latest transport measurement is
that the superconductivity in Tl and K intercalated FeSe materials
is proximity to an AFM insulating phase \cite{fang}, similar as in
high-T$_c$ cuprates. In particular, TlFe$_x$Se$_2$ ($1.3<x<1.7$) is
found to be an AFM insulator, with an activated transport gap of
$\sim 57$ meV for $x=1.5$ \cite{fang}. This has revived the
discussion on the correlation effect in Fe-based superconductors.

To clarify this issue, we have performed the first-principles
electronic structure calculations on TlFe$_{1.5}$Se$_2$ and
AFe$_{1.5}$Se$_2$ (A=K, Rb, or Cs). We find that the ground state of
crystal TlFe$_{1.5}$Se$_2$ is indeed an AFM semiconductor with a
collinear AFM order (Fig. 4(b)) and an energy gap of 94 meV, and
similarly AFe$_{1.5}$Se$_2$ is also an AFM semiconductor but with a
zero-gap semiconducting state or AFM semimetallic state close to the
ground state. This is the first time theoretically to show there
exists an AFM insulating state in Fe-based superconductor materials.

In our calculations the plane wave basis method was used
\cite{pwscf}. We adopted the generalized gradient approximation
(GGA) with Perdew-Burke-Ernzerhof formula \cite{pbe} for the
exchange-correlation potentials. The ultrasoft pseudopotentials
\cite{vanderbilt} were used to model the electron-ion interactions.
After the full convergency test, the kinetic energy cut-off and the
charge density cut-off of the plane wave basis were chosen to be 800
eV and 6400 eV, respectively. The Gaussian broadening technique was
used and a mesh of $18\times 18\times 9$ k-points were sampled for
the Brillouin-zone integration. In the calculations, the lattice
parameters with the internal atomic coordinates were optimized by
the energy minimization. For TlFe$_{1.5}$Se$_2$, the optimized
tetragonal lattice parameters are found in excellent agreement with
the experimental ones \cite{fang}.

\begin{figure}
\includegraphics[width=8.0cm]{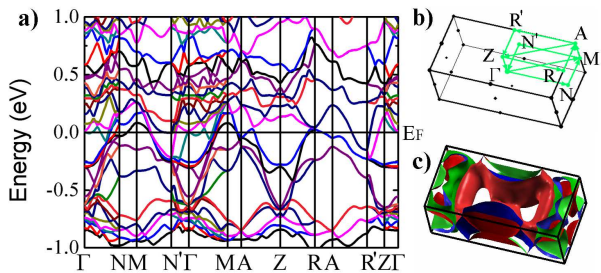}
\caption{(Color online) Electronic structure of TlFe$_{1.5}$Se$_2$
with the Fe-vacancies ordered in rhombus (Fig. 1b) in the
nonmagnetic state: (a) the band structure,  (b) the Brillouin zone,
and (c) the Fermi surface. The Fermi energy sets to zero.
}\label{figb}
\end{figure}

\begin{figure}
\includegraphics[width=8.0cm]{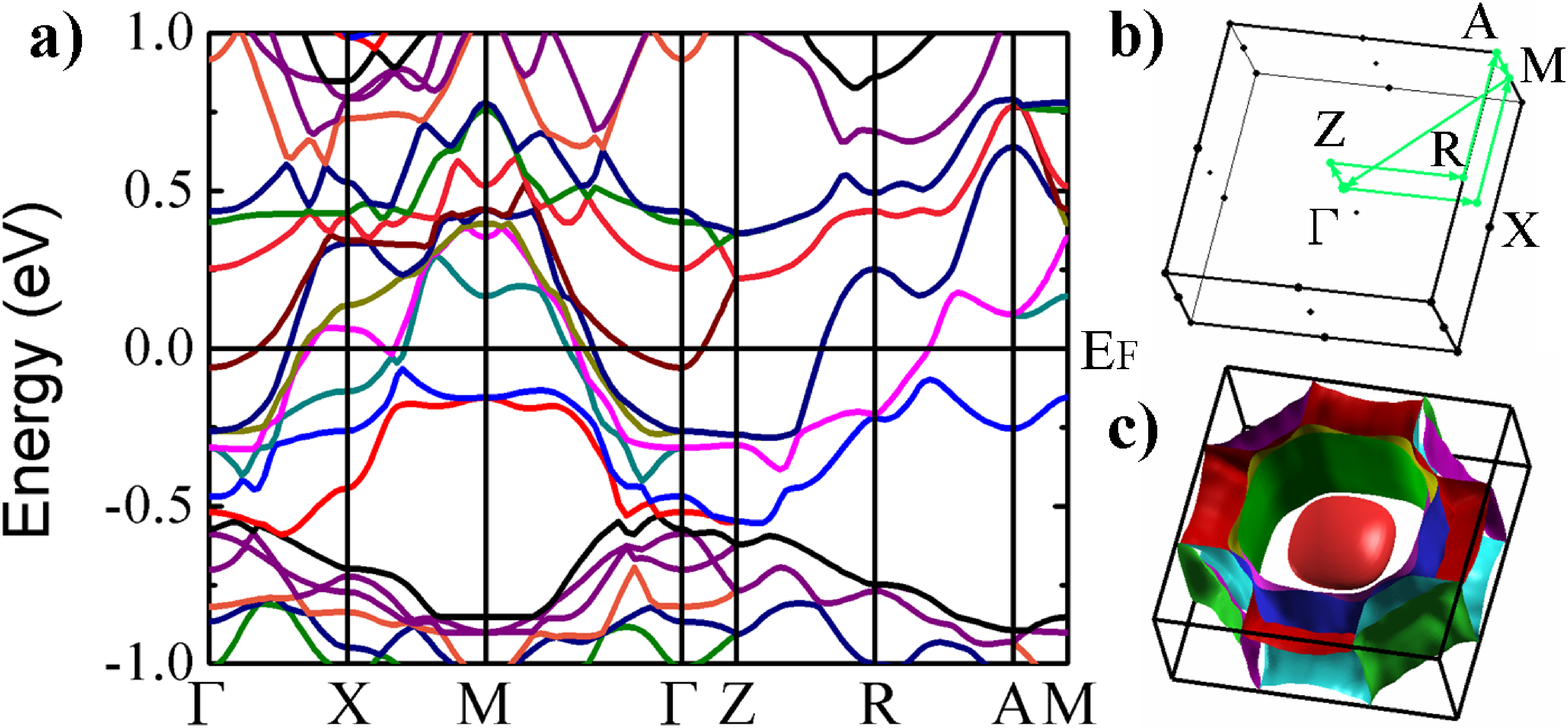}
\caption{(Color online) Electronic structure of TlFe$_{1.5}$Se$_2$
with the Fe-vacancies ordered in square (Fig. 1c) in the nonmagnetic
sate: (a) the band structure, (b) the Brillouin zone, and (c) the
Fermi surface. The Fermi energy sets to zero.} \label{figc}
\end{figure}

The early experiments suggested that the Fe-vacancies in
TlFe$_{1.5}$Se$_2$ are ordered in a rhombus structure as shown in
Fig. 1(b)\cite{sab}. In order to test theoretically whether this
structure is truly the ground state, we have also calculated the
electronic structure of another possible one-quarter Fe-vacancies
ordered structure, namely the square-ordered vacancy structure as
shown in Fig. 1(c). In the nonmagnetic state, we find that the
energy of the square-ordered vacancy structure is lower by about 12
meV/Fe than the rhombus-ordered vacancy structure. It is noted that
an Fe-vacancy only induces small local structural deviation (less
than 0.04 \AA ) from the original tetragonal one except for Tl
atoms, which means the covalent bonding between Fe and Se atoms is
rather robust. Figs. 2 and 3 show the nonmagnetic band structures
and Fermi surfaces of TlFe$_{1.5}$Se$_2$ in the Fe-vacancies
rhombus- and square-ordered structures, respectively. In both cases,
there are a number of electron and hole bands crossing the Fermi
level. The Fermi surface nesting is not as strong as in the iron
pnictides \cite{lu1}.

However, in the true ground state which is in an AFM order, we find
that the energy of the rhombus-ordered vacancy structure is lower by
15.1 meV/Fe than the square-ordered one. Figs. 4 and 5 show a number
of possible magnetic orders in the Fe-vacancies rhombus- and
square-ordered structures, respectively. Among these magnetic
ordered states, we find that the Fe-vacancies rhombus-ordered
structure with an A-collinear AFM order shown in Fig. 4(b) has the
lowest energy. This rhombus order of vacancies agrees with the
neutron measurement\cite{sab}. Our result further suggests that the
rhombus-ordered vacancy structure is stabilized by an A-collinear
AFM order. In the A-collinear AFM state (Fig. 4(b)), the Fe moments
are antiferromagnetically ordered along the lines without Fe
vacancies and ferromagnetically ordered along the lines
perpendicular.

\begin{figure}
\includegraphics[width=7cm]{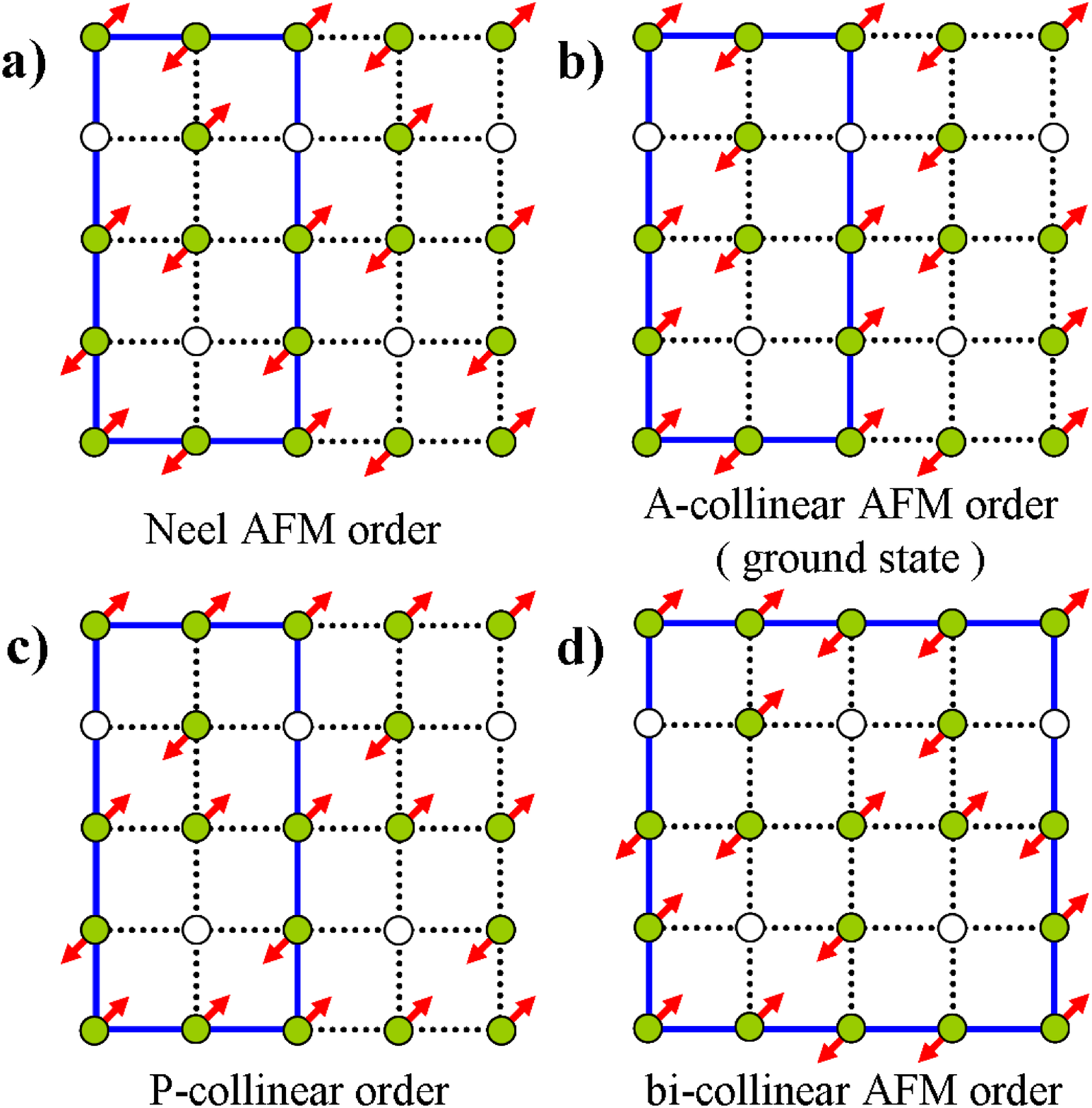}
\caption{(Color online) Schematic top view of four possible magnetic
orders in the Fe-Fe square layer with one quarter Fe-vacancies
ordered in rhombus: (a) checkboard Neel order in which the nearest
neighboring Fe moments are anti-parallel ordered; (b) A-collinear
AFM order in which the Fe moments are antiferromangtic ordered along
the line without vacancies; (c) P-collinear AFM order in Fe moments
are antiferromagnetic ordered along the lines with vacancies; (d)
bi-collinear AFM order. The squares or rectangles enclosed by the
solid lines denote the magnetic unit cells. }\label{figd}
\end{figure}

\begin{figure}
\includegraphics[width=7cm]{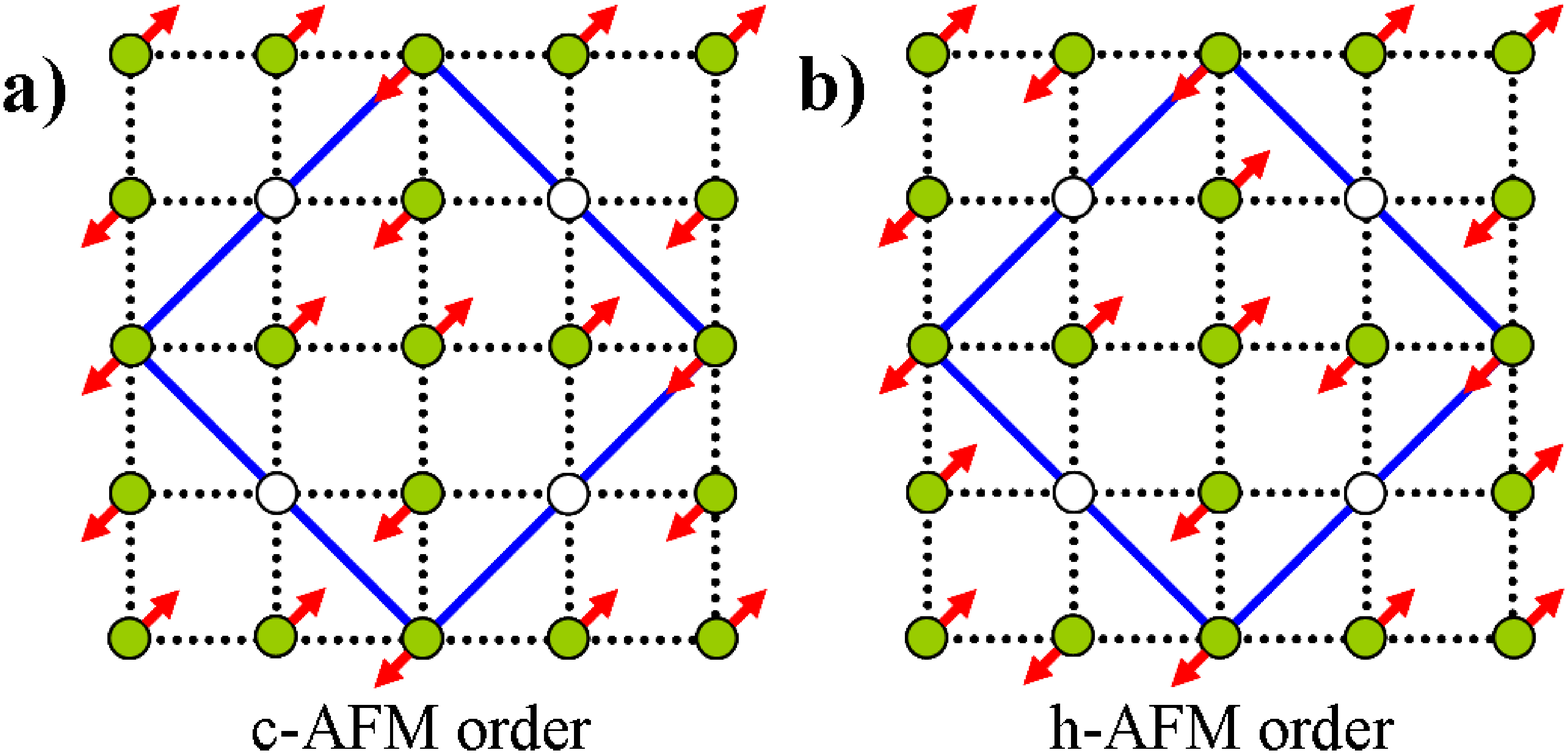}
\caption{(Color online) Schematic top view of two possible magnetic
orders in the Fe-Fe square layer: (a) c-AFM order in which the next
nearest Fe moments are anti-parallel ordered; (b) h-AFM order in
which half pairs of the next nearest Fe moments in anti-parallel
while the other half pairs in parallel. The squares enclosed by the
solid lines denote the magnetic unit cells. }\label{fige}
\end{figure}

Similar as in the iron-pnictides\cite{ma1,ma2}, we find that there
is a small structural distortion in TlFe$_{1.5}$Se$_2$. The lattice
constant slightly expands along the AFM direction and contracts
along the ferromagnetic direction in the A-collinear AFM state. This
leads to a small energy gain of $\sim$1 meV/Fe. We also find that
the Fe magnetic moments between the neighbor layers FeSe are
antiferromagnetically ordered. The energy difference between the
ferromagnetic and AFM interlayer magnetic states is $\sim$4.9
meV/Fe. Thus the ground state of TlFe$_{1.5}$Se$_2$ is an
A-collinear AFM state with AFM interlayers of FeSe.

\begin{figure}
\includegraphics[width=7.5cm]{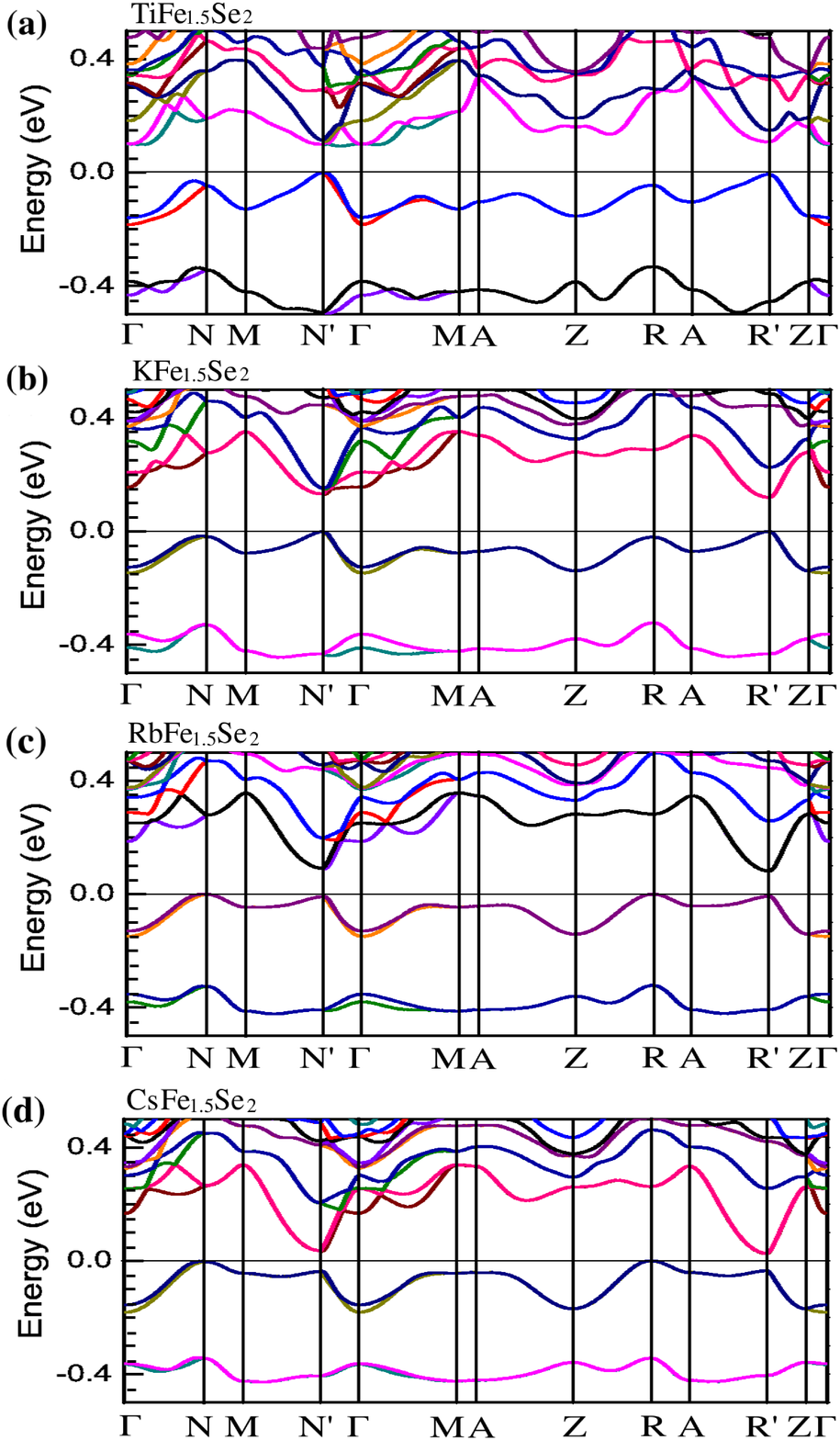}
\caption{(Color online) Electronic band structure of
TlFe$_{1.5}$Se$_2$ or AFe$_{1.5}$Se$_2$ (A=K, Rb, or Cs) with one
quarter Fe-vacancies ordered in rhombus (see Fig. 1(b)) and with an
A-collinear AFM order (see Fig. 4 (b)) in the ground state, which is
an antiferromagnetic semiconductor. (a) TlFe$_{1.5}$Se$_2$; (b)
KFe$_{1.5}$Se$_2$; (c) RbFe$_{1.5}$Se$_2$; (d) CsFe$_{1.5}$Se$_2$.
The Brillouin zone is shown in Fig. 2(b). Here the top of the
valence band sets to zero. Note that
$\Gamma$-N($\Gamma$-N$^{\prime}$) corresponds to the
anti-parallel(parallel)-aligned moment line. } \label{figf}
\end{figure}

Fig. 6(a) shows the ground state electronic band structure of
TlFe$_{1.5}$Se$_2$. Unlike other parent compounds of iron pnictides
or chalcogenides, TlFe$_{1.5}$Se$_2$ is an AFM semiconductor, in
agreement with experimental measurements. The energy band gap is
found to be $\sim 94$ meV, which is also consistent with the gap
value obtained by the transport measurement, 57.7 meV \cite{fang}.
The compound TlFe$_{1.5}$Se$_2$ in the other magnetic orders is
found mostly to be metallic.

\begin{figure}
\includegraphics[width=7.0cm]{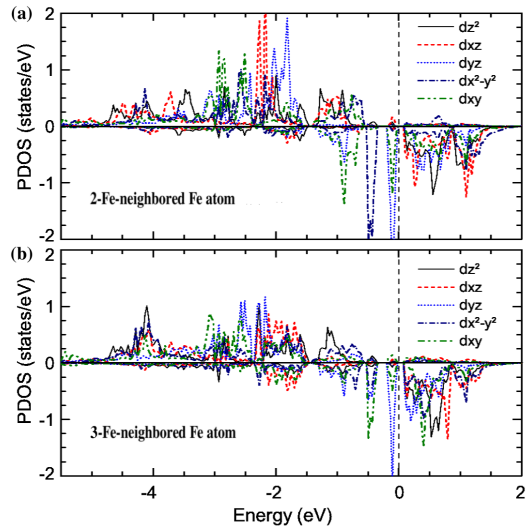}
\caption{(Color online) Projected density of states at the five
Fe-$3d$ orbitals around (a) 2-Fe-neighbored Fe atom and (b)
3-Fe-neighbored atom in the A-collinear AFM ordered ground state of
TlFe$_{1.5}$Se$_2$. Here the top of the valence band sets to zero.
And $x$-axis is along the antiferromagnetic direction while $y$-axis
is along the ferromagnetic direction in Fig. 4(b). }\label{figd}
\end{figure}

There are two inequivalent Fe atoms according to the number of
neighboring Fe atoms if the Fe-vacancies are rhombus-ordered,namely
2-Fe-neighbored and 3-Fe-neighbored Fe atoms respectively (see Fig.
1(b)). By projecting the density of states onto the five $3d$
orbitals of Fe (Fig. 7), we find that the five up-spin orbitals are
almost completely filled on both kinds of Fe atoms. This suggests
that the crystal field splitting induced by Se atoms is small,
similar as in iron-pnictides. The magnetic moment formed around each
Fe atom is found to be about 2.8$\mu_B$. This large Fe magnetic
moments apparently results from the Hund's rule
coupling\cite{ma1,ma2}. The down spin orbitals are partially filled
by $d_{yz}$, $d_{xy}$, and $d_{x^2-y^2}$ orbitals for the
2-Fe-neighbored Fe atoms or by $d_{yz}$ and $d_{xy}$ orbitals for
the 3-Fe-neighbored Fe atoms. Such anisotropy among the five down
spin orbitals in TlFe$_{1.5}$Se$_2$ results from the Fe-vacancies.

Inspection of the real space charge distribution in
TlFe$_{1.5}$Se$_2$ shows that there is a strong covalence bond
between the neighbor Fe and Se atoms. This gives rise to an
effective AFM superexchange interactions bridged by the Se atoms
between the next nearest neighbor Fe moments, similar as in iron
pnictides \cite{ma1,ma2}. If the energy of the nonmagnetic state is
set to zero, we find that the energies of the ferromagnetic,
checkboard Neel AFM, A-collinear AFM, and bi-collinear AFM states
are (-0.114, -0.217, -0.326, -0.270) eV/Fe for TlFe$_{1.5}$Se$_2$.
The magnetic moment around each Fe atom is found to be about
$2.5$-$2.8\mu_B$, varying weakly in the above four magnetically
ordered states, similar as in the iron-pnictides \cite{ma1,ma2}. If
we attribute the energy differences of these states entirely to the
contribution of magnetic interactions between the spins $\vec{S}$
and model them by the simple Heisenberg model with the nearest
($J_1)$, next-nearest ($J_2$), and next-next nearest ($J_3$)
neighbor interactions and ignore the anisotropy caused by the Fe
vacancies \cite{ma1}, we find that $J_1=39~meV/S^2$,
$J_2=60~meV/S^2$, and $J_3=7~meV/S^2$ for TlFe$_{1.5}$Se$_2$. In
obtaining these values, the contribution from itinerant electrons is
ignored. Similarly we find that $J_1=37~meV/S^2$, $J_2=64~meV/S^2$,
and $J_3=8~meV/S^2$ for KFe$_{1.5}$Se$_2$ (see below for more).

To quantify the electronic correlation effect in TlFe$_{1.5}$Se$_2$,
we have performed a GGA+$U$ calculation. We find that the energy
band gap increases dramatically with the Coulomb interaction $U$.
When $U$ is larger than 2~eV, the energy band gap is over 190~meV,
which is significantly larger than the measurement value of 57
meV\cite{fang}. This suggests that the magnetic ordering and the
energy band gap are driven mainly by the exchange effect, rather
than the correlation effect.

Here we emphasize that the calculation convergency test needs to be
elaborated. An enough high energy cutoff and a set of sufficient
many k-points are required to ensure the correct ground state,
namely an AFM semiconductor obtained. Otherwise, the calculations
always yield a metallic state rather than semiconducting state.

We have further performed the first-principles electronic structure
calculations for AFe$_{1.5}$Se$_2$ (A=K, Rb, or Cs). Like
TlFe$_{1.5}$Se$_2$, we find that crystal AFe$_{1.5}$Se$_2$ is an AFM
semiconductor and its ground state is also A-collinear AFM state
with AFM interlayers of FeSe. Figs. 6 (b), (c), and (d) show the
ground state electronic band structures for KFe$_{1.5}$Se$_2$,
RbFe$_{1.5}$Se$_2$, and CsFe$_{1.5}$Se$_2$, respectively.
Accordingly, the energy band gaps are found to be 121, 69, and 26
meV for KFe$_{1.5}$Se$_2$, RbFe$_{1.5}$Se$_2$, and
CsFe$_{1.5}$Se$_2$, respectively.

The calculations show that the energies of KFe$_{1.5}$Se$_2$,
RbFe$_{1.5}$Se$_2$, and CsFe$_{1.5}$Se$_2$ in the A-collinear AFM
order but with the ferromagnetic interlayers of FeSe are higher than
the AFM interlayers of FeSe by 8.9, 5.2, and 2.8 meV/Fe,
respectively. In such a case with the ferromagnetic interlayers of
FeSe, crystals TlFe$_{1.5}$Se$_2$ and KFe$_{1.5}$Se$_2$ both are
still AFM semiconductors with indirect band gaps of 20 and 32 meV
and direct gaps of 46 and 62 meV, respectively. However, crystal
RbFe$_{1.5}$Se$_2$ becomes a zero-gap semiconductor while crystal
CsFe$_{1.5}$Se$_2$ becomes an AFM semimetal with the electron and
hole carrier densities of $7.47\times 10^{19}/cm^3$ and $7.35\times
10^{19}/cm^3$, respectively. These two states, zero-gap
semiconducting and AFM semimetallic, will significantly influence
the physical properties of RbFe$_{1.5}$Se$_2$ and CsFe$_{1.5}$Se$_2$
since they are almost degenerated with the ground states.

Moreover, we also find that the square-ordered vacancy structure
with a c-AFM order (Fig. 5(a)) and AFM interlayers of FeSe are
metastable for crystal AFe$_{1.5}$Se$_2$ (A=K, Rb, or Cs).
Energetically, it is just higher than the ground state by 10.9, 7.9,
and 5.0 meV/Fe for KFe$_{1.5}$Se$_2$, RbFe$_{1.5}$Se$_2$, and
CsFe$_{1.5}$Se$_2$, respectively. Such a square-ordered vacancy
structure pattern should thus have a certain probability of
formation in the compound AFe$_{1.5}$Se$_2$.


In conclusion, we have performed the first principles calculations
for the electronic structure and magnetic order of
TlFe$_{1.5}$Se$_2$. We find that the crystal structure of
TlFe$_{1.5}$Se$_2$ with the rhombus-ordered Fe vacancies in a
collinear antiferromagnetic order is energetically the most stable.
The ground state of TlFe$_{1.5}$Se$_2$ is a quasi-two-dimensional
antiferromagnetic semiconductor with an energy gap of 94 meV. Our
result is consistent with both neutron and transport measurements.
We further predict that crystal AFe$_{1.5}$Se$_2$ (A=K, Rb, or Cs)
is an antiferromagnetic semiconductor like TlFe$_{1.5}$Se$_2$, and
there is a zero-gap semiconducting state or antiferromagnetic
semimetallic state nearly degenerated with the ground state for
RbFe$_{1.5}$Se$_2$ or CsFe$_{1.5}$Se$_2$.

This work is partially supported by National Natural Science
Foundation of China and by National Program for Basic Research of
MOST, China.

\end{document}